\documentclass[12pt, titlepage]{article}
\usepackage{authblk}
\usepackage[margin=1in]{geometry}
\RequirePackage{amsthm,amsmath,amsfonts}
\RequirePackage{natbib}
\RequirePackage[colorlinks,citecolor=blue,urlcolor=blue]{hyperref}
\usepackage{booktabs,longtable}
\usepackage{setspace}
\usepackage{graphicx}
\usepackage{xcolor}
\usepackage{epstopdf}
\usepackage{tikz}
\usetikzlibrary{arrows.meta}
\usepackage{subcaption}
\graphicspath{{./}{../figures/}}
\usepackage{bm}
\usepackage[ruled, lined]{algorithm2e}

\allowdisplaybreaks

\newcommand{\ta}{a}
\newcommand{\tb}{b}
\newcommand{\bZ}{\bm{Z}}
\newcommand{\outs}{s^{\, (1)}}
\newcommand{\ins}{s^{\, (2)}}
\newcommand{\targetW}{\bm{\Lambda}}
\newcommand{\diffW}{\bm{\Psi}}
\newtheorem{proposition}{Proposition}[section]
\newcommand{\red}[1]{\textcolor{red}{#1}}
\newcommand{\blue}[1]{\textcolor{blue}{#1}}
\newcommand{\pkg}[1]{{\fontshape{it}\selectfont #1}}

\newcommand{\proglang}[1]{\textsf{#1}}

\title{A Strength and Sparsity Preserving Algorithm for Generating 
	Weighted, Directed Networks with Predetermined Assortativity}

\author[1]{Yelie Yuan}

\author[1]{Jun Yan}

\author[2,*]{Panpan Zhang}

\affil[1]{Department of Statistics, University of Connecticut, 
	Storrs, CT 06269}

\affil[2]{Department of Biostatistics, Vanderbilt University Medical 
	Center, Nashville, TN 37203}

\affil[*]{Correspondence: 
	\href{mailto:panpan.zhang@vumc.org}{panpan.zhang@vumc.org}}

\begin{document}
	
	\maketitle
	
	\begin{abstract}
		Degree-preserving rewiring is a widely used technique for 
		generating 
		unweighted networks with given assortativity, but for 
		weighted
		networks, it is unclear how an analog would preserve the 
		strengths and
		other critical network features such as sparsity level. This 
		study
		introduces a novel approach for rewiring weighted networks 
		to achieve
		desired directed assortativity. The method utilizes a mixed 
		integer
		programming framework to establish a target network with
		predetermined assortativity coefficients, followed by an 
		efficient
		rewiring algorithm termed ``strength and sparsity preserving 
		rewiring''
		(SSPR). SSPR retains the node strength distributions and 
		network
		sparsity after rewiring. It is also possible to accommodate 
		additional
		properties like edge weight distribution with extra 
		computational
		cost. The optimization scheme can be used to determine 
		feasible
		assortativity ranges for an initial network. The 
		effectiveness of the
		proposed SSPR algorithm is demonstrated through its 
		application to
		two classes of popular network models.

		\bigskip
		
		
	\end{abstract}
	
	\doublespacing
	
	\section{Introduction}
	\label{sec:intro}

	Assortativity is an important measure characterizing 
	the correlation structure of nodal features of networks. 
	Generating
	random networks with predetermined assortativity is critical for 
	justifying network theories~\citep{Newman2003mixing}, exploring 
	spectral properties~\citep{VanMieghem2010influence}, improving 
	model fit~\citep{Wang2022generating}, and optimizing network 
	robustness~\citep{Chen2016optimizing}. Edge rewiring is a widely 
	accepted technique for generating networks with given 
	assortativity. The degree-preserving 
	rewiring (DPR) algorithm proposed by~\citet{Newman2003mixing} 
	ensures that the node degree distribution of an undirected 
	network 
	keeps unchanged throughout the course of rewiring so as to 
	preserve 
	the fundamental topology of the rewired network. An extension of 
	Newman's algorithm, called DiDPR, was recently developed for 
	generating directed networks with predetermined directed 
	assortativity coefficients~\citep{Wang2022generating}. More 
	generally, rewiring techniques have 
	found practical applications in many fields such as biological
	science~\citep{Iorio2016efficient}, clinical
	trials~\citep{Staples2015incorporating}, and social network 
	analysis \citep{Schweimer2022generating}, among others.

	Despite the long availability of Newman's algorithm for
	unweighted networks, rewiring a weighted network to achieve
	predetermined assortativity proves to be a challenging task that 
	has
	not yet been studied in the literature. While directly extending
	Newman's algorithm to weighted networks may seem feasible for
	preserving node strengths, a naive extension fails to retain 
	other
	important network properties, such as sparsity, simultaneously.
	When edge weights are represented as integer values, one 
	potential
	approach involves a multi-edge scheme, where a weighted edge is
	divided into multiple unit-weight edges, and then Newman's 
	two-swap
	method or similar techniques are applied. However, this approach 
	does
	not translate smoothly to networks with real-valued edge 
	weights. 
	Even for networks with integer-valued edge weights, this method 
	lacks
	practical utility, as it results in a substantial increase in the
	number of edges~\citep{Wuellner2010resilience} and alters the
	network's sparsity, a typical feature observed in most real-world
	networks~\citep{Barabasi2016network}. In the subsequent section, 
	we
	will delve into more details about this approach and explore 
	other
	potential possibilities for addressing the challenge of rewiring
	weighted networks to achieve predetermined assortativity.

	This paper introduces a novel rewiring algorithm for generating
	weighted, directed networks with four predetermined directed
	assortativity coefficients~\citep{Yuan2021assortativity}. 
	Notably,
	the proposed algorithm ensures that both out- and in-strength 
	distributions, along with sparsity, are meticulously preserved 
	upon the completion of the rewiring process. Newman's approach 
	involves searching for a target network structure characterized 
	by
	a joint degree distribution that matches the desired 
	assortativity
	values and represents the stationary status of the rewiring 
	process~\citep{Newman2003mixing, Wang2022generating}. In 
	contrast, 
	the proposed algorithm directly produces a weighted, directed 
	network
	with assortativity measures precisely equal to the given values,
	provided that they are attainable. The desired network structure 
	is
	not unique and we formulate an optimization problem to provide 
	one
	feasible solution.
	With carefully chosen objective function for the optimization, 
	the
	algorithm can retain certain network topology and properties in
	addition to strength distributions and sparsity after rewiring.
	Further, the optimization scheme also helps to identify
	the attainability of each assortativity coefficient by 
	establishing
	its upper and lower bounds given the initial network 
	configuration.

	The remainder of the paper is organized as follows.
	Section~\ref{sec:preliminary} introduces the notations and 
	elucidates the challenges for the extension of Newman's 
	algorithm to 
	weighted networks. Section~\ref{sec:sspr} presents an efficient 
	strength and sparsity preserving reserving algorithm for 
	weighted, 
	directed networks with given assortativity coefficients, 
	followed by 
	an approach to determining assortativity coefficient bounds and 
	a 
	generalization allowing to consider other network properties 
	like 
	edge weight distribution. Section~\ref{sec:simulation} provides 
	extensive simulations showing the applications of the proposed 
	algorithm to the Erd\"{o}s-R\'{e}nyi model and the 
	Barab\'{a}si-Albert model. Lastly, some discussions and future 
	works 
	are addressed in Section~\ref{sec:discussion}.

	\section{Preliminaries}
	\label{sec:preliminary}
	
	Starting with notations, we layout the challenges when extending 
	Newman's algorithm to weighted, directed networks.
	
	\subsection{Notations}
	\label{sec:notations}
	
	Let $G := G(V, E)$ be a weighted, directed network with node 
	set~$V$ 
	and edge set~$E$. Additionally, let $(v_i, v_j, w_{ij}) \in E$ 
	denote a weighted, directed edge from source node $v_i \in V$ to 
	target node $v_j \in V$ with weight $w_{ij} > 0$. For the 
	special 
	case of $v_i = v_j$,
	$(v_i, v_j, w_{ij}) \in E$ is a self-loop. Network $G$ is 
	characterized by its associated adjacency matrix
	$\bm{W} :=  (w_{ij})_{n \times n}$,
	where $n = |V|$ is the number of nodes in~$G$.
	If there is no edge from~$v_i$ to $v_j$, i.e.,
	$(v_i, v_j, w_{ij}) \notin E$,
	then the corresponding $w_{ij}$ in $\bm{W}$ is set to~$0$.
	Fundamental node-level properties of weighted, directed 
	networks are $\outs_{i} := \sum_{v_j \in V} w_{ij}$ and
	$\ins_{i} := \sum_{v_j \in V} w_{ji}$,
	which respectively refer to the out- and 
	in-strength of node $v_i$. The superscripts ``$1$'' and ``$2$'' 
	are respectively used to represent ``out'' and ``in'' throughout 
	the 
	rest of the manuscript for simplicity.

	The directed assortativity coefficients considered in this paper 
	are adopted from those proposed by 
	\citet{Yuan2021assortativity}. By 
	considering the combinations of out- and in-strengths of source 
	and 
	target nodes, there are four types of assortativity 
	coefficients, 
	denoted by $r(\ta, \tb) = r_{\bm{W}}(\ta, \tb)$ with 
	$\ta, \tb \in \{1, 2\}$, where $r(\ta, \tb)$ is the
	assortativity coefficient based on the $\ta$-strength of source
	nodes and $\tb$-strength of target nodes.
	For example, $r(1, 2)$ refers to the 
	assortativity coefficient based on the out-strength of source 
	nodes 
	and in-strength of target nodes. The rest three are interpreted 
	in 
	the similar manner.

	Mathematically, the directed assortativity coefficients 
	are expressed as
	\begin{equation}
		\label{eq:fourd}
		\displaystyle
		r(\ta, \tb) = \frac{
			\sum_{v_i, v_j \in V} w_{ij}
			\left[\left(s_i^{(\ta)} - \bar{s}_{\rm src}^{\, (\ta)} 
			\right)
			\left(s_j^{(\tb)} - \bar{s}_{\rm trg}^{\, (\tb)} 
			\right)\right]
		}{
			\tau \sigma_{\rm src}^{(\ta)} \sigma_{\rm trg}^{(\tb)}
		},
		\qquad
		\ta, \tb \in \left\{1, 2\right\},
	\end{equation}
	where $\tau := \sum_{v_i, v_j \in V}w_{ij}$ is the total weight 
	of all edges,
	\begin{equation*}
		\bar{s}_{\rm src}^{\, (\ta)} := 
		\frac{\sum_{v_i, v_j \in V} w_{ij} s_i^{\, (\ta)}}{\tau} = 
		\frac{\sum_{v_i \in V} \outs_i s_i^{\, (\ta)}}{\tau}
	\end{equation*}
	is the weighted mean of the $\ta$-type strength of source nodes 
	and
	\begin{equation*}
		\sigma_{\rm src}^{\, (\ta)} := 
		\sqrt{\frac{\sum_{v_i, v_j \in V} w_{ij} \left(s_i^{\, 
		(\ta)} - 
				\bar{s}_{\rm src}^{\, (\ta)}\right)^2}{\tau}} =	
		\sqrt{\frac{\sum_{v_i \in V} \outs_i \left(s_i^{\, (\ta)} - 
				\bar{s}_{\rm src}^{\, (\ta)}\right)^2}{\tau}}
	\end{equation*}
	is the associated weighted standard deviation. The counterparts 
	$\bar{s}_{\rm trg}^{\, (\tb)}$ and $\sigma_{\rm trg}^{\, (\tb)}$ 
	are defined analogously for target nodes. For more properties 
	about
	the directed, weighted assortativity coefficients, see
	\citet{Yuan2021assortativity}.

	\subsection{Challenges}
	\label{sec:challenges}
	
\begin{figure}
		\begin{subfigure}[t]{0.5\linewidth}
			\centering
			\begin{tikzpicture}
				\draw (0, 2) node[draw = black, circle, minimum size 
				= 
				0.75cm] (v1) {$v_1$};
				\draw (2, 2) node[draw = black, circle, minimum size 
				= 
				0.75cm] (v2) {$v_2$};
				\draw (0, 0) node[draw = black, circle, minimum size 
				= 
				0.75cm] (v3) {$v_3$};
				\draw (2, 0) node[draw = black, circle, minimum size 
				= 
				0.75cm] (v4) {$v_4$};
				\draw[-latex] (v1) -- (v3);
				\draw[-latex] (v2) -- (v4);
				\draw[-{Triangle[width = 5mm, length = 3mm]}, line 
				width 
				= 
				2mm] (2.5, 1) -- (3.5, 1);
				\draw (4, 2) node[draw = black, circle, minimum size 
				= 
				0.75cm] (v5) {$v_1$};
				\draw (6, 2) node[draw = black, circle, minimum size 
				= 
				0.75cm] (v6) {$v_2$};
				\draw (4, 0) node[draw = black, circle, minimum size 
				= 
				0.75cm] (v7) {$v_3$};
				\draw (6, 0) node[draw = black, circle, minimum size 
				= 
				0.75cm] (v8) {$v_4$};
				\draw[-latex, dashed] (v5) -- (v7);
				\draw[-latex, dashed] (v6) -- (v8);
				\draw[-latex] (v5) -- (v8);
				\draw[-latex] (v6) -- (v7);
			\end{tikzpicture}
		\end{subfigure}
		\hfill
		\begin{subfigure}[t]{0.5\linewidth}
			\centering
			\begin{tikzpicture}
				\draw (0, 2) node[draw = black, circle, minimum size 
				= 
				0.75cm] (v1) {$v_1$};
				\draw (2, 2) node[draw = black, circle, minimum size 
				= 
				0.75cm] (v2) {$v_2$};
				\draw (0, 0) node[draw = black, circle, minimum size 
				= 
				0.75cm] (v3) {$v_3$};
				\draw (2, 0) node[draw = black, circle, minimum size 
				= 
				0.75cm] (v4) {$v_4$};
				\draw[-latex, blue] (v1) -- node [midway, right, 
				blue] 
				{3} (v3);
				\draw[-latex, red] (v2) -- node [midway, left, red] 
				{5} 
				(v4);
				\draw[-{Triangle[width = 5mm, length = 3mm]}, line 
				width 
				= 
				2mm] (2.5, 1) -- (3.5, 1);
				\draw (4, 2) node[draw = black, circle, minimum size 
				= 
				0.75cm] (v5) {$v_1$};
				\draw (6, 2) node[draw = black, circle, minimum size 
				= 
				0.75cm] (v6) {$v_2$};
				\draw (4, 0) node[draw = black, circle, minimum size 
				= 
				0.75cm] (v7) {$v_3$};
				\draw (6, 0) node[draw = black, circle, minimum size 
				= 
				0.75cm] (v8) {$v_4$};
				\draw[-latex, dashed] (v5) -- (v7);
				\draw[-latex, red] (v6) -- node [midway, right, red] 
				{2} 
				(v8);
				\draw[-latex, blue] (v5) -- node [near end, above, 
				blue] 
				{3} (v8);
				\draw[-latex, red] (v6) -- node [near end, above, 
				red] 
				{3} (v7);
			\end{tikzpicture}
		\end{subfigure}
		\begin{center}
			\begin{subfigure}[t]{0.5\linewidth}
				\centering
				\begin{tikzpicture}
					\draw (0, 2) node[draw = black, circle, minimum 
					size = 
					0.75cm] (v1) {$v_1$};
					\draw (2, 2) node[draw = black, circle, minimum 
					size = 
					0.75cm] (v2) {$v_2$};
					\draw (0, 0) node[draw = black, circle, minimum 
					size = 
					0.75cm] (v3) {$v_3$};
					\draw (2, 0) node[draw = black, circle, minimum 
					size = 
					0.75cm] (v4) {$v_4$};
					\draw[-latex, blue] (v1) -- node [midway, right, 
					blue] 
					{3} (v3);
					\draw[-latex, orange] (v1) -- node [midway, 
					above, 
					orange] {2} (v4);
					\draw[-latex, red] (v2) -- node [midway, left, 
					red] {5} 
					(v4);
					\draw[-{Triangle[width = 5mm, length = 3mm]}, 
					line width 
					= 2mm] (2.5, 1) -- (3.5, 1);
					\draw (4, 2) node[draw = black, circle, minimum 
					size = 
					0.75cm] (v5) {$v_1$};
					\draw (6, 2) node[draw = black, circle, minimum 
					size = 
					0.75cm] (v6) {$v_2$};
					\draw (4, 0) node[draw = black, circle, minimum 
					size = 
					0.75cm] (v7) {$v_3$};
					\draw (6, 0) node[draw = black, circle, minimum 
					size = 
					0.75cm] (v8) {$v_4$};
					\draw[-latex, dashed] (v5) -- (v7);
					\draw[-latex, orange] (v6) -- node [midway, 
					left, 
					orange] {2} (v8);
					\draw[-latex, red] (v5) -- node [near end, 
					above, red] 
					{5} (v8);
					\draw[-latex, blue] (v6) -- node [near end, 
					above, blue] 
					{3} (v7);
				\end{tikzpicture}
			\end{subfigure}
		\end{center}
		\caption{Top-left panel: An illustration of DPR in 
			\cite{Newman2003mixing} and DiDPR in 
			\cite{Wang2022generating}; 
			Dashed edges refer to the edges that are removed in the 
			rewiring; Each edge has weight $1$. Top-right panel: An 
			illustration of DPR directly extended to weighted, 
			directed 
			networks; Edge weights are given next to the 
			corresponding 
			edges. Bottom panel: An illustration of three-swap 
			rewiring for 
			weighted, directed networks.}
		\label{fig:dpr}
	\end{figure}
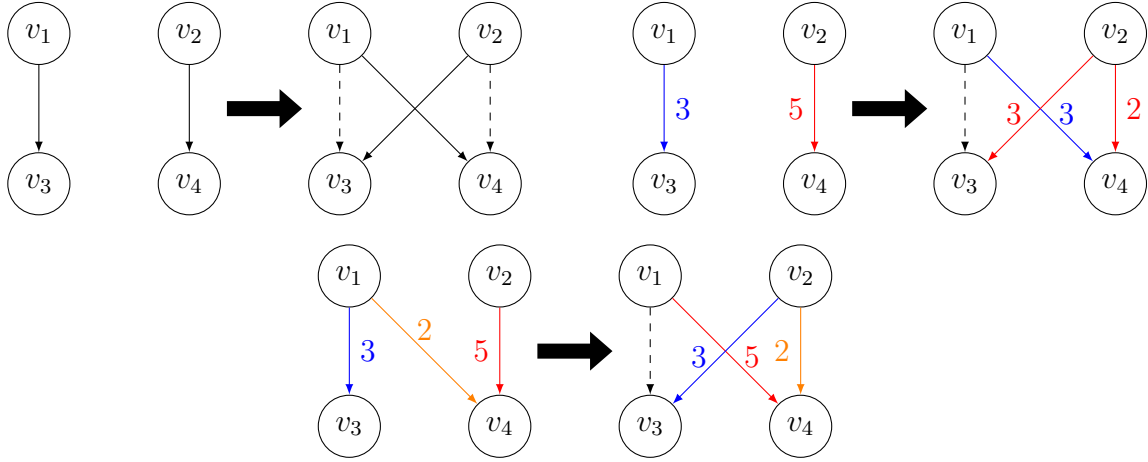

	Newman's algorithm for generating an unweighted, 
	undirected network with a predetermined assortativity measure is 
	based on a two-swap DPR algorithm~\citep{Newman2003mixing}. It
	effectively adjusts the assortativity while preserving the
	marginal node degree distribution. Recently, this idea was 
	translated
	into a practical approach with concrete via a convex optimization
	framework, and further extended to
	unweighted, directed networks by~\citet{Wang2022generating}. The 
	extension solely requires accounting for edge directions during 
	the
	rewiring process, and since all edges possess unit weight, both 
	node
	degrees and the total edge number remain unchanged. For a 
	graphical
	illustration, refer to the top-left panel of 
	Figure~\ref{fig:dpr}.
	
	An example that attempts to extend Newman's algorithm to a 
	weighted, 
	directed network~\citep{Wuellner2010resilience} is shown in the 
	top-right panel of Figure~\ref{fig:dpr}. As depicted, however, 
	when
	the sampled edges for swap have different weights, 
	an additional edge with weight equal to their weight difference 
	needs
	to be added to preserve node strengths;
	the rightmost red edge (with weight $2$) from $v_2$ 
	to $v_4$ illustrates this necessity. Consequently, this attempt
	becomes impractical, as it leads to a significant increase in the
	number of edges during the rewiring process, thereby compromising
	network
	sparsity. For certain special cases, a potential remedy is to
	generalize the three-swap idea introduced 
	by~\citet{Uribe2020finding}.
	The bottom panel of Figure~\ref{fig:dpr} demonstrates an example 
	of
	this approach, which is applicable only to simple edge weights 
	(e.g.,
	integer-valued) and demands additional efforts to search 
	specific 
	structures for rewiring to occur. In the provided example, 
	it requires sampling a module of four nodes connected by three 
	directed edges, where the out-strengths of the source nodes 
	(i.e., 
	$v_1$ and $v_2$) are identical. Candidate structures satisfying 
	such
	restrictions may be scarce or non-existent, making methods
	based on three-swap impractical in real-world scenarios.

	Furthermore, an issue not investigated by 
	\citet{Newman2003mixing} is
	whether the predetermined assortativity level is achievable 
	through
	rewiring. Newman's algorithm requires the development of a
	transition matrix $\bm{M}$ to construct
	the joint edge degree distribution for the target network with
	predetermined assortativity. The existence of this matrix
	$\bm{M}$ is not guaranteed as it depends on both the structure of
	the initial network and the predetermined assortativity value. In
	other words, given an initial network, not every
	predetermined assortativity level is attainable through rewiring.
	Inspired by the work of \citet{Wang2022generating}, as a 
	byproduct,
	this paper also delves into an investigation of
	assortativity attainability by determining upper and lower 
	bounds for
	each of the four directed assortativity coefficients, 
	conditional on 
	the structure of the initial network.

	\section{Strength and Sparsity Preserving Rewiring}
	\label{sec:sspr}
	
	We propose an efficient strength and sparsity preserving rewiring
	(SSPR) algorithm designed for weighted, directed networks with
	predetermined assortativity coefficients. The crux of the 
	algorithm
	lies in the quest for a target network with the desired 
	assortativity
	coefficients by solving a mixed integer linear programming
	problem. Subsequently, the algorithm employs a novel rewiring
	technique to ensure the preservation of critical network 
	properties,
	such as marginal strength distributions and network sparsity.
	the rewiring process.
	
	\subsection{Finding a Target Network}
	\label{sec:targetW}
	
	Given a fully observed network $G := G(V(G), E(G))$ with a 
	weighted 
	adjacency matrix $\bm{W}$ and predetermined assortativity 
	measures 
	$r^*(\ta, \tb), \ta, \tb \in \left\{1, 2\right\}$, the primary
	goal is to generate a new network $H$ (defined on the same node 
	set 
	$V(H) = V(G) = V$, but with a different edge set $E(H) \neq 
	E(G)$) 
	whose assortativity measures are equal to the given
	$r^*(\ta, \tb)$
	through rewiring $G$. Meanwhile, it is essential to retain
	crucial network properties like node out- and in-strength 
	distributions and network sparsity after rewiring. 
	Provided that such $H$ exists, its adjacency matrix
	$\targetW := (\lambda_{ij})_{n \times n}$,
	which is referred to as the target adjacency matrix, must 
	satisfy 
	the following conditions:
	\begin{enumerate}
		\item[(1)] The entries of $\targetW$ are non-negative, i.e.,
		$\lambda_{ij} \ge 0$ for all $v_i, v_j \in V$;
		\item[(2)] The row and column sums of $\targetW$ are 
		identical to 
		the counterparts in $\bm{W}$ (preserving marginal strength 
		distributions);
		\item[(3)] The number of non-zero elements in $\targetW$ is 
		the 
		same as that in $\bm{W}$ (preserving network sparsity);
		\item[(4)] The assortativity measures (of $\targetW$) 
		computed 
		from Equation~\eqref{eq:fourd} are equal to the given 
		$r^*(\ta, \tb)$ for all $\ta, \tb \in \{1, 2\}$.
	\end{enumerate}
	Depending on the analytic objectives and computing resources, 
	one may
	also include additional conditions that restrict the
	lower and upper bounds of the non-zero elements in $\targetW$ in 
	order
	to prevent the emergence of a large proportion of extremely 
	small 
	weights or unexpected outliers.

	Now the problem boils down to finding a suitable target adjacency
	matrix~$\targetW$, which may not be unique, but any single 
	solution
	would suffice. To set it up, consider a latent, binary matrix 
	$\bZ := (z_{ij})_{n \times n}$ associated with $\targetW$, where
	$z_{ij} = 1$ for $\lambda_{ij} > 0$; $z_{ij} = 0$ otherwise. The
	search for a solution to $\targetW$ involves solving a mixed 
	integer
	linear programming problem as follows:
	\begin{align*}
		\min_{\targetW} \quad &f(\targetW), \nonumber\\
		\textrm{s.t.}  \quad 
		&\lambda_{ij} = 0 \textrm{ if } z_{ij} = 0, 
		\quad \forall v_i, v_j \in V, \\
		&\kappa_{\rm L} \le \lambda_{ij} \le \kappa_{\rm U} \textrm{ 
		if 
		} z_{ij} = 1, \quad \forall v_i, v_j \in V, \\ 
		&\sum_{i, j} z_{ij} = \vert E(G) \vert, \\
		&\sum_{i} \lambda_{ij} = \sum_{i} w_{ij} = \ins_{j}, 
		\quad \forall v_j \in V,  \\
		&\sum_{j} \lambda_{ij}  = \sum_{j} w_{ij} = \outs_i, 
		\quad \forall v_i \in V, \\
		&r_{\targetW}(\ta, \tb) = r^* (\ta, \tb),
		\quad \ta, \tb \in \{1, 2\},
	\end{align*}
	where $\kappa_{\rm U} \ge \kappa_{\rm L} > 0$ are the preset 
	upper and lower
	bounds for edge weights, and $f(\cdot)$ is an arbitrary linear
	function.

	In theory, the objective function $f(\cdot)$ can be any function
	of~$\targetW$ and the constraints do not have to be linear
	in~$\targetW$. For instance, one may set 
	$f(\targetW) = \sum_{i, j} \vert w_{ij} - \lambda_{ij} \vert$ 
	if it is desired that the edge 
	weight distribution changes as little as possible after 
	rewiring. 
	Nonetheless, the more complex the objective function is and the 
	more 
	additional constraints are, the more time is required for 
	solving 
	the optimization problem even with a possibility of unsolvable 
	risks. Especially when there are non-linear constraints, the 
	optimization problem becomes a mixed integer 
	non-linear programming problem, which demands more solving time 
	or even computationally intractable. Therefore, when there is no 
	mandatory condition for $f(\cdot)$, we recommend setting it to 
	zero
	for improving the optimization speed. Mathematical programming 
	solvers like Gurobi~\citep{gurobi} and CPLEX~\citep{cplex} can 
	be 
	used to efficiently solve such problems.

	A byproduct of the optimization scheme is that it can be used to
	determine the bounds of feasible assortativity levels.
	Given an initial network $G$, not all the values in the 
	natural bounds of assortativity coefficient (i.e., $[-1, 1]$) 
	are 
	attainable through SSPR that will be elaborated in the next 
	subsection. From Equation~\eqref{eq:fourd}, 
	the assortativity coefficients are linear in edge weights, 
	allowing 
	us to find the assortativity bounds by adjusting the objective 
	function. Specifically, we can set the objective function to be
	$f(\targetW) = \sum_{i, j} \lambda_{ij} s_i^{(\ta)} s_j^{(\tb)}$ 
	to find the lower bound of $r^*(\ta, \tb)$, and set 
	$f(\targetW) = - \sum_{i, j} \lambda_{ij} s_i^{(\ta)} 
	s_j^{(\tb)}$ 
	to find the upper bound of $r^*(\ta, \tb)$. See detailed 
	illustrations in Section~\ref{sec:simulation}.

	\subsection{Rewiring towards Target Network}
	\label{sec:rewirepath}
	
\begin{figure}[tbp]
		\begin{subfigure}{\linewidth}
			\centering
			\begin{tikzpicture}[scale=0.75]
				\draw (-7, 2) node[draw=black, circle, minimum 
				size=0.8cm] (i) 
				{$v_i$};
				\draw (-3, 2) node[draw=black, circle, minimum 
				size=0.8cm] (j) 
				{$v_j$};
				\draw (-7, -2) node[draw=black, circle, minimum 
				size=0.8cm] 
				(l) {$v_l$};
				\draw (-3, -2) node[draw=black, circle, minimum 
				size=0.8cm] 
				(k) {$v_k$};
				\draw[-latex, thick] (i) -- node [midway, above] 
				{$w_{ij}$} (j);
				\draw[-latex, thick] (k) -- node [midway, below] 
				{$w_{kl}$} (l);
				\draw[-latex, thick, dotted] (i) -- node [midway, 
				below, 
				sloped] 
				{$w_{il}$} (l);
				\draw[-latex, thick, dotted] (k) -- node 
				[midway, above, sloped, rotate=180] {$w_{kj}$} (j);
				
				\draw (3, 2) node[draw=black, circle, minimum 
				size=0.8cm] (i2) 
				{$v_i$};
				\draw (7, 2) node[draw=black, circle, minimum 
				size=0.8cm] (j2) 
				{$v_j$};
				\draw (3, -2) node[draw=black, circle, minimum 
				size=0.8cm] 
				(l2) {$v_l$};
				\draw (7, -2) node[draw=black, circle, minimum 
				size=0.8cm] 
				(k2) {$v_k$};
				
				\draw[-latex, thick, dotted] (i2) -- node [midway, 
				above] 
				{$w_{ij} - \Delta w$} (j2);
				\draw[-latex, thick, dotted] (k2) -- node [midway, 
				below] 
				{$w_{kl} - \Delta w$} (l2);
				\draw[-latex, thick] (i2) -- node [midway, below, 
				sloped] 
				{$w_{il} + \Delta w$} (l2);
				\draw[-latex, thick] (k2) -- node [midway, above, 
				sloped] 
				{$w_{kj} + \Delta w$} (j2);
				
			\end{tikzpicture}
		\end{subfigure}
		\begin{subfigure}{\linewidth}
			\centering
			\begin{equation*}
				\begin{pmatrix}
					\cdots & \vdots & \cdots & \vdots & \cdots \\
					\cdots & w_{ij} & \cdots & w_{il} & \cdots \\
					\cdots & \vdots & \cdots & \vdots & \cdots \\
					\cdots & w_{kj} & \cdots & w_{kl} & \cdots \\
					\cdots & \vdots & \cdots & \vdots & \cdots \\
				\end{pmatrix}
				\Longrightarrow
				\begin{pmatrix}
					\cdots & \vdots     & \cdots & \vdots     & 
					\cdots \\
					\cdots & w_{ij} - \Delta w & \cdots & w_{il} + 
					\Delta w & 
					\cdots \\
					\cdots & \vdots     & \cdots & \vdots     & 
					\cdots \\
					\cdots & w_{kj} + \Delta w & \cdots & w_{kl} - 
					\Delta w & 
					\cdots \\
					\cdots & \vdots     & \cdots & \vdots     & 
					\cdots \\
				\end{pmatrix}
			\end{equation*}
		\end{subfigure}
		\caption{A hypothetical example reflecting the principle of 
			rewiring through a graphical representation and the 
			corresponding 
			changes in the adjacency matrix.}
		\label{fig:rewire}
	\end{figure}
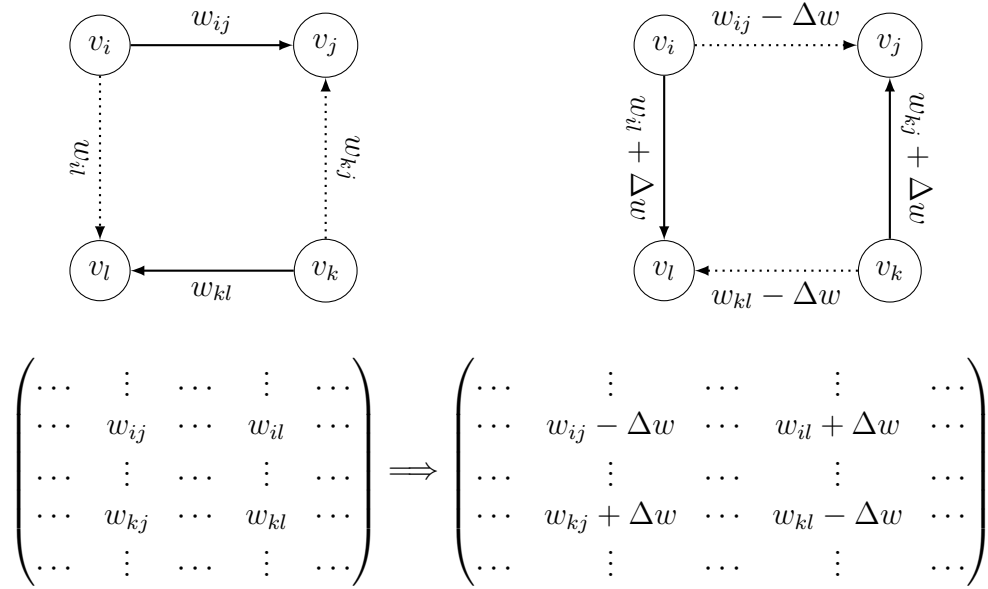

	Once the target adjacency matrix~$\targetW$ is determined, the 
	next 
	crucial task is to establish a feasible rewiring scheme to move 
	from
	given $\bm{W}$ towards $\targetW$ while preserving node in- and 
	out-strengths and network sparsity. 
	Figure~\ref{fig:rewire} shows a 
	hypothetical example of rewiring a pair of edges 
	$(v_i, v_j, w_{ij})$ and $(v_k, v_l, w_{kl})$ among four
	nodes $v_i$, $v_j$, $v_k$ and $v_l$. The underlying 
	principle is to keep the out- and in-strengths of the four 
	nodes identical by a meticulously redistributed weight of amount
	$\Delta w \le \min\{w_{ij}, w_{kl}\}$. It is worth
	noting that the directed edges $(v_k, v_j, w_{kj})$ and
	$(v_i, v_l, w_{il})$ may not exist before rewiring, and that
	the selected edges $(v_i, v_j, w_{ij})$ and $(v_k, v_l, w_{kl})$
	may be removed after rewiring, so represented by dotted lines.
	The corresponding changes in the adjacency matrix are 
	illustrated in
	the lower panel of Figure~\ref{fig:rewire}. It is clear that the 
	row
	and column sums in the adjacency matrix remain unchanged.
	An appropriate $\Delta w$ must be determined for each rewiring 
	step. 
	
	Before proceeding, however, it is crucial to show the existence
	of at least one rewiring path from~$\bm{W}$ to~$\targetW$. To 
	achieve
	this, define
	$\diffW := (\psi_{ij})_{n \times n} = \bm{W} - \targetW$, the
	difference between the initial adjacency matrix~$\bm{W}$ and the
	target adjacency matrix~$\targetW$. A successful rewiring process
	means that, at the end of the rewiring, $\psi_{ij} = 0$ for all
	$i, j \in \{1, \ldots, n\}$. We employ a sweeping procedure to 
	adjust all $\psi_{ij}$'s one by one in an order from the top to 
	the 
	bottom and the left to the right within each row. The existence 
	of a
	path is shown by induction.
	Suppose that $\psi_{ij}$ is the next element in the sweeping 
	procedure
	awaiting an adjustment via rewiring. That is, we already have
	$\psi_{kl} = 0$ for all $l$ if $k < i$ and for $l < j$ if $k = i$
	from previous sweeping steps. Proposition~\ref{prop:exist}
	shows the existence of a rewiring path leading to $\psi_{ij} = 
	0$ 
	with the associated proof given in 
	Appendix~\ref{app:proof_exist}.

	\begin{proposition}
		\label{prop:exist}
		For any $i,j < n$, there always exists a path leading to 
		$\psi_{ij} = 0$ after rewiring.
	\end{proposition}

	\begin{figure}[tbp]
		\centering
		\begin{equation*}
			\begin{pmatrix}
				0 & 0 & 0 & 0 \\
				0 & \blue{3} & -2 & -1 \\
				1 & -1 & \red{2} & -2 \\
				-1 & -2 & 0 & \red{3} 
			\end{pmatrix} \overset{\Delta w = 2}{\Longrightarrow}
			\begin{pmatrix}
				0 & 0 & 0 & 0 \\
				0 & \blue{1} & 0 & -1 \\
				1 & 1 & \red{0} & -2 \\
				-1 & -2 & 0 & \red{3} 
			\end{pmatrix} \overset{\Delta w = 1}{\Longrightarrow}
			\begin{pmatrix}
				0 & 0 & 0 & 0 \\
				0 & \blue{0} & 0 & 0 \\
				1 & 1 & \red{0} & -2 \\
				-1 & -1 & 0 & \red{2} 
			\end{pmatrix}.
		\end{equation*}
		\caption{An example of rewiring scheme illustrating 
			Proposition~\ref{prop:exist}.}
		\label{fig:prop}
	\end{figure}
	
	Figure~\ref{fig:prop} illustrates an example of a rewiring scheme
	corresponding to the principle behind 
	Proposition~\ref{prop:exist}.
	The objective is to transform $\psi_{22} = 3$ (shown in blue) 
	to~$0$
	through rewiring. The $\psi_{i^\ast j^\ast}$'s to be
	adjusted are highlighted in red, with the corresponding $\Delta 
	w$
	values indicated above the arrows. Notably, the selection of
	$\Delta w$ values is not unique. To reduce subsequent rewiring
	steps, we prioritize making non-sweeped $\psi_{ij}$'s zero 
	whenever
	possible. Specifically, for each pair $(k, l)$ with $k > i$ and 
	$l > 
	j$, given
	$\psi_{ij} > 0$, we set
	$\Delta w = \min\{\psi_{ij}, \max\{\psi_{kl}, 0\}\}$;
	given $\psi_{ij} < 0$, we set
	$\Delta w = \max\{\psi_{ij}, \min\{-\psi_{kj}, 0\}, 
	\min\{-\psi_{il}, 0\}\}$.
	These additional 
	conditions for $\Delta w$ selection are essential to prevent the
	generation of negative edge weights during the rewiring process.

	\begin{algorithm}[tbp]
		\caption{Pseudo codes of the SSRP algorithm.}
		\label{alg:rewire}
		\SetNoFillComment
		\KwIn{Initial adjacency matrix $\bm{W}$; \newline
			target adjacency matrix $\targetW$.}
		\KwOut{Rewiring record $R$.}
		\SetKwProg{Algr}{Algorithm}{:}{}
		\SetKwProg{Fn}{Function}{:}{}
		\SetKwFunction{Rewire}{Rewire}
		\SetKwFunction{Reorder}{Reorder}
		\Algr{}{
			$n \leftarrow$ number of rows (or columns) of $\bm{W}$\;
			Initialize an empty list of rewiring steps $R$\;
			$\diffW \leftarrow \bm{W} - \targetW$\;
			\For{$i = 1$ \KwTo $n - 1$} {
				\tcc{Possibly insert a reorder step here to put 
				elements with larger magnitude earlier}
				\For{$j = 1$ \KwTo $n - 1$} {
					$\diffW, R\leftarrow$ \Rewire{$\diffW, R, i, j, 
					n$}\;
				}
			}
			\KwRet{$R$}\;
		}
		
		\hrulefill
		
		\Fn{\Rewire}{
			\KwIn{Matrix $\diffW$; \newline
				rewiring record $R$; \newline
				row and column indices $i$ and $j$; \newline
				number of rows $n$.}
			\KwOut{Updated $\diffW$ and $R$.}
			\For{$k = i + 1$ \KwTo $n$} {
				\For{$l = j + 1$ \KwTo $n$}{
					\uIf{$\psi_{ij} > 0$} {
						$\Delta w \leftarrow \min(\psi_{ij}, \max(0, 
						\psi_{k, l}))$\;
					}
					\uElseIf {$\psi_{ij} < 0$} {
						$\Delta w \leftarrow \max(\psi_{ij}, 
						\min(0, -\psi_{i, l}), \min(0, -\psi_{k, 
						j}))$\;
					}
					\uIf{$\psi_{ij} == 0$ \rm{\textbf{or}} $\Delta w 
					== 0$}{
						Continue to the next $l$\;
					}
					$\psi_{i j} \leftarrow \psi_{i j} - \Delta w$
					\tcc*{Update $\diffW$}
					$\psi_{k l} \leftarrow \psi_{k l} - \Delta w$\;
					$\psi_{i l} \leftarrow \psi_{i l} + \Delta w$\;
					$\psi_{k j} \leftarrow \psi_{k j} + \Delta w$\;
					\uIf(\tcc*[f]{Record rewiring step}){$\Delta w > 
					0$} {
						Append 
						$(i, j, k, l, \Delta w)$ 
						to $R$\;
					}
					\uElse{
						Append 
						$(i, l, k, j, -\Delta w)$ 
						to $R$\;
					}
				}
			}
			\KwRet{$\diffW$, $R$}\;
		}
	\end{algorithm}

	The pseudo codes for the SSPR algorithm are summarized in
	Algorithm~\ref{alg:rewire}. Given the difference 
	matrix~$\diffW$, 
	the \texttt{Rewire} function sweeps through its elements one 
	at a time. Note that the sweep only needs to be done for the 
	first
	$n - 1$ rows as the column sums are zero; similarly,
	within each row of $\diffW$, we only need sweep the first 
	$n - 1$ elements as all the row sums are zero. At the beginning 
	of
	each row, an optional step is to reorder the rows and columns so 
	that
	elements with larger magnitude get sweeped earlier. This would 
	reduce
	the number of rewiring steps (about 45\% in our experiments in
	Section~\ref{sec:simulation}), but the extra sorting
	step would increase the time complexity of the algorithm. The 
	output
	of Algorithm~\ref{alg:rewire} is the list of entire rewiring 
	history
	of $(i, j, k, l, \Delta w)$ for each step, where 
	$i$, $j$, $k$, and $l$ are the indices of the selected nodes 
	$v_i$, $v_j$, $v_k$, and $v_l$ for rewiring, and $\Delta w$ is 
	the 
	associated rewiring weight.

	\section{Simulations}
	\label{sec:simulation}
	
	We validate the proposed SSPR algorithm through 
	simulation studies using two widely used network models: the 
	Erd\"{o}s-R\'{e}nyi (ER) model~\citep{Erdos1959on, 
		Gilbert1959random} and the Barab\'{a}si-Albert 
	model, also known as the preferential attachment (PA)
	model~\citep{Barabasi1999emergence}. Both models in their 
	classic 
	forms are unweighted, but, in our study, they are extended by 
	incorporating edge directions and weights. The algorithm 
	implementation is primarily based on the 
	\pkg{gurobipy} module~\citep{gurobi} in \proglang{Python}, and 
	the 
	program was run on AMD EPYC 7763 processors utilizing 4 threads 
	and 
	8 GB of memory.

	\subsection{ER Network Model}
	\label{sec:ER}
	
	The classic ER model is governed by two parameters: 
	the number of nodes $n$ and the probability of 
	emergence of a directed edge~$p$. We augment the classic ER 
	model by 
	allowing self-loops (from a node to itself) and edge weights. 
	Specifically, three levels of
	$n \in \{50, 100, 200\}$ and three levels of
	$p \in \{0.05, 0.1, 0.2\}$ were considered. Edge weights were
	generated from a gamma distribution with shape~$5$ and
	scale~$0.2$. For each configuration, a total of $100$ ER 
	networks 
	were generated. Isolated nodes, if any, were
	removed from the network prior to rewiring. Pertaining to the 
	nature 
	of the ER model, all of the four assortativity coefficients are 
	expected to converge to~$0$ for large~$n$.

	\begin{figure}[tbp]
		\centering
		\includegraphics[width=0.75\textwidth]{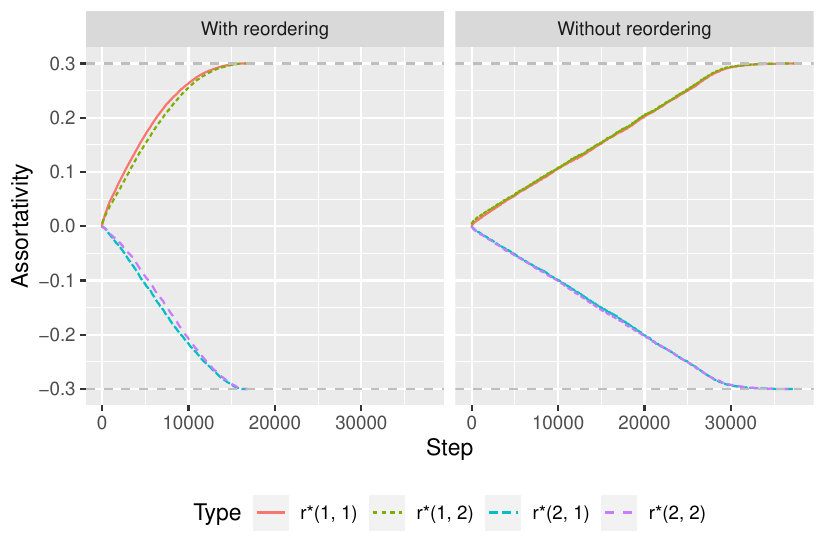}
		\caption{Average trace plots for $100$ replicates of ER 
		networks 
			with $n = 200$, $p = 0.1$, target assortativity 
			coefficients set 
			to $r^*(1, 1) = r^*(1, 2) = 0.3$, $r^*(2, 1) = r^*(2, 2) 
			= -0.3$, 
			and objective function given by $f(\targetW) = 0$. The 
			left panel 
			shows the results with reordering, but the right panel 
			shows those 
			without reordering.}
		\label{fig:ER_f0}
	\end{figure}

	Figure~\ref{fig:ER_f0} shows the results for $n = 200$ and $p = 
	0.1$
	as an example. The results for other settings of $n$ and $p$ 
	present 
	a similar pattern, so they are omitted. The target assortativity
	coefficients were $r^*(1, 1) = r^*(1, 2) = 0.3$, 
	$r^*(2, 1) = r^*(2, 2) = -0.3$, and a simple objective function
	$f(\targetW) = 0$ was used to determine the target adjacency 
	matrix
	for the assortativity coefficients. The left panel presents the 
	results with the reordering procedure implemented, and the right 
	panel presents the results without reordering. Each
	panel shows the average trace plots for the four assortativity 
	coefficients during rewiring. We observe that all
	of the assortativity coefficients successfully reached their 
	targets
	through the proposed SSPR algorithm, regardless of whether the
	reordering procedure was implemented. However, the right panel 
	shows 
	a significant increase in the number of rewiring steps when the
	reordering procedure was not executed into the SSPR algorithm.

	\begin{figure}[tbp]
		\centering
		\begin{subfigure}[t]{0.63\textwidth}
			\vskip 0pt
			\includegraphics[width = \textwidth]
			{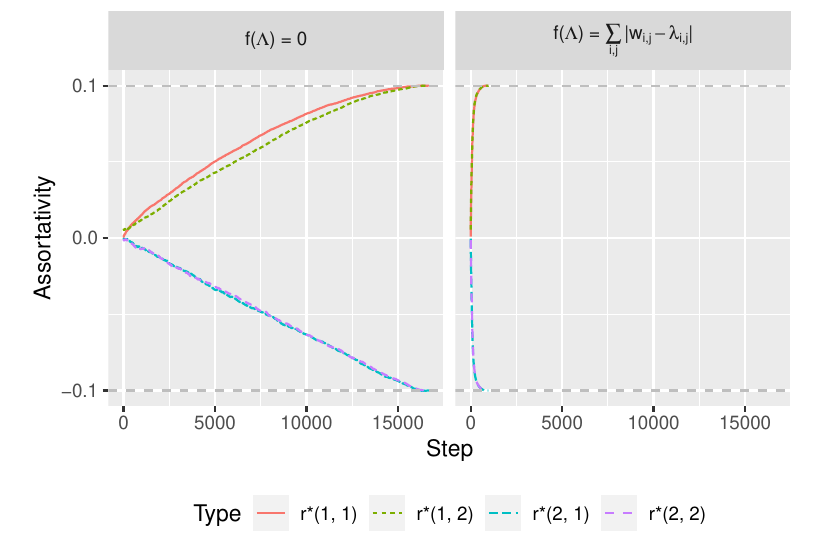}
		\end{subfigure}
		\begin{subfigure}[t]{0.36\textwidth}
			\vskip 0pt
			\includegraphics[width = 
			\textwidth]{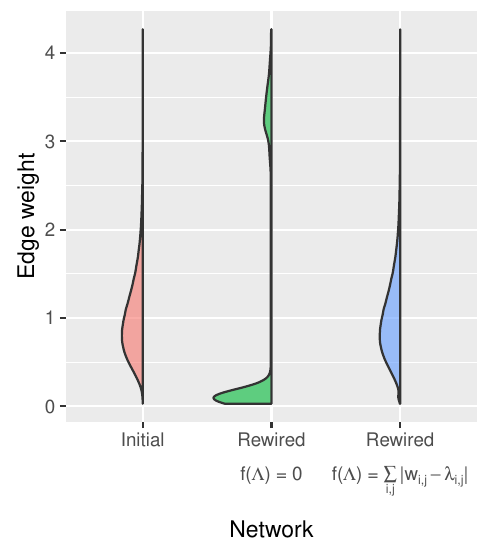}
		\end{subfigure}
		\caption{Average trace and edge weight violin plots 
			for $100$ replicates of ER networks with $n = 200$ and 
			$p = 0.1$,
			target assortativity coefficients set to
			$r^*(1, 1) = r^*(1, 2) = 0.1$ and $r^*(2, 1) = r^*(2, 2) 
			= 
			-0.1$.}
		\label{fig:ER_f1}
	\end{figure}

	To illustrate the impact of the selection of objective function
	$f(\cdot)$, consider rewiring ER networks with $n = 200$
	and $p = 0.1$ to achieve assortativity coefficients
	$r^*(1, 1) = r^*(1, 2) = 0.1$ and $r^*(2, 1) = r^*(2, 2) = -0.1$.
	Two objective functions $f(\targetW) = 0$ and 
	$f(\targetW) = \sum_{i,j} |w_{ij} - \lambda_{ij}|$ were used in
	setting up the mixed integer programming
	problem in Section~\ref{sec:targetW}.
	The average trace
	plots of the assortativity levels based on $100$ simulated 
	networks 
	are displayed in the left two panels of Figure~\ref{fig:ER_f1}.
	Clearly, the algorithm with the more complex objective function 
	$f(\targetW) = \sum_{i,j} |w_{ij} - \lambda_{ij}|$
	requires much fewer rewiring steps for the assortativity 
	coefficients to reach the predetermined targets. The right panel 
	of 
	Figure~\ref{fig:ER_f1} compares the density of the edge weights 
	of the
	initial networks and the rewired networks obtained
	under the two objective functions. The post-rewiring edge weight
	distribution with objective function $f(\targetW) = 0$ is 
	noticeably
	different from that of the initial networks. In contrast, the
	post-rewiring edge weight distribution with objective function 
	$f(\targetW) = \sum_{i,j} |w_{ij} - \lambda_{ij}|$ is almost 
	identical to that of the initial networks.

	The comparisons in Figure~\ref{fig:ER_f1} seems suggesting
	a preference of using 
	$f(\targetW) = \sum_{i,j} |w_{ij} - \lambda_{ij}|$ 
	as the objective function, but there are other factors to 
	consider.
	Fewer rewiring steps do not necessarily mean less overall 
	computation
	time. In fact, among the $100$ replicates in the present 
	example, 
	the median computation time for $f(\targetW) = 0$ was $5$ 
	seconds, 
	but for $f(\targetW) = \sum_{i,j} |w_{ij} - \lambda_{ij}|$ it 
	was 
	about $191$ seconds. Further, there is no guarantee that 
	optimization
	problem with the more complex 
	objective function can be solved within a reasonable amount of 
	time. 
	For instance, for the experiment of ER networks with $n = 100$ 
	and 
	$p = 0.01$, $48$ out of the $100$ simulations did not finish
	within $12$ hours on a computer with AMD 
	EPYC 7763 processors with 4~threads and 8~GB of RAM.

	\begin{figure}[tbp]
		\centering
		\includegraphics[width = 0.75\textwidth]{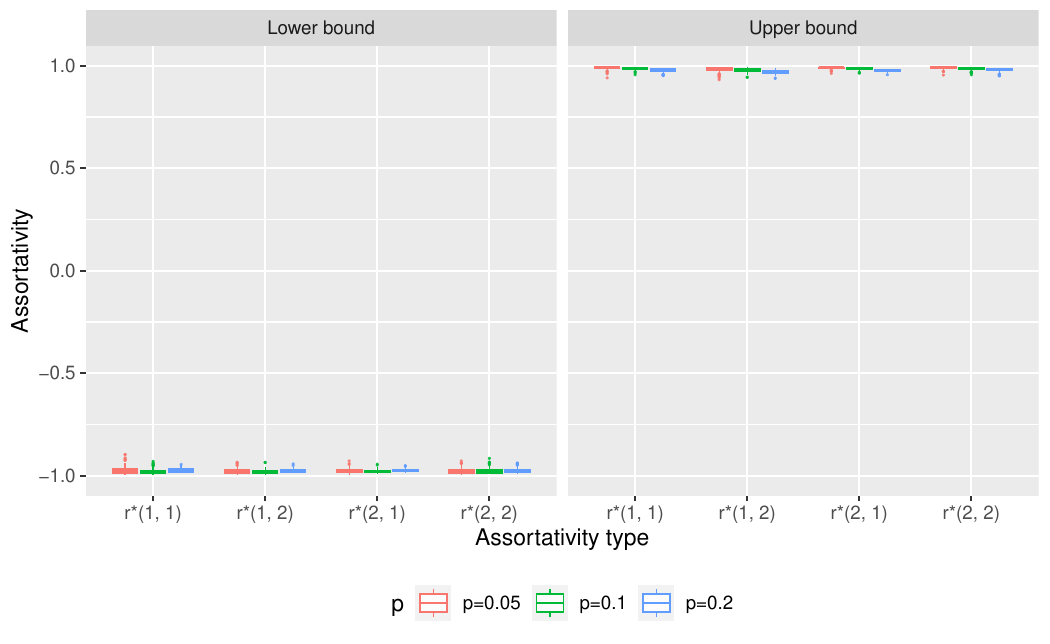}
		\caption{Side-by-side box plots for the upper and lower 
		bounds of
			assortativity coefficients based on $100$ replicates of 
			ER 
			networks with $n = 200$ and $p \in \{0.05, 0.1, 0.2\}$.}
		\label{fig:ER_bound}
	\end{figure}

	Figure~\ref{fig:ER_bound} shows the box plots of
	the attainable upper and lower bounds of assortativity 
	coefficients 
	for the 100 ER networks generated under each combination of $n = 
	200$
	and $p \in \{0.05, 0.1, 0.2\}$. It appears that the bounds are 
	very
	close to the nominal bounds of $-1$ and $1$. That is,
	all the values between $-1$ and $1$ appear to be attainable for 
	all
	four assortativity coefficients. Such observation is expected 
	owing 
	to the feature of ER networks.

	\subsection{PA Network Model}
	\label{sec:PA}

	\begin{figure}[h]
		\begin{center}
			\begin{tikzpicture}
				\draw[fill = gray!20] (-5, 0) ellipse (1.6 and 1) ;
				\draw (-4.2, 0) node[draw = black, circle, minimum
				size =
				0.8cm, fill = blue!20] (j1) {$v_j$} ;
				\draw (-3.5, -1.5) node[draw = black, circle,
				minimum
				size = 0.8cm] (i1)  {$v_i$} ;
				\draw[-latex, thick] (i1) -- (j1) ;
				\draw[fill = gray!20] (0, 0) ellipse (1.6 and 1) ;
				\draw (-0.8, 0) node[draw = black, circle, minimum
				size =
				0.8cm, fill = blue!20] (i2) {$v_i$} ;
				\draw (0.8, 0) node[draw = black, circle, minimum
				size = 0.8cm, fill = blue!20] (j2)  {$v_j$} ;
				\draw[-latex, thick] (i2) -- (j2) ;
				\draw[fill = gray!20] (5, 0) ellipse (1.6 and 1) ;
				\draw (4.2, 0) node[draw = black, circle, minimum
				size =
				0.8cm, fill = blue!20] (i3) {$v_i$} ;
				\draw (3.5, -1.5) node[draw = black, circle, minimum
				size = 0.8cm] (j3)  {$v_j$} ;
				\draw[-latex, thick] (i3) -- (j3) ;
			\end{tikzpicture}
			\caption{The $\alpha$, $\beta$ and $\gamma$ 
				edge-creation scenarios (from left to right).}
			\label{fig:PAscenario}
		\end{center}
	\end{figure}
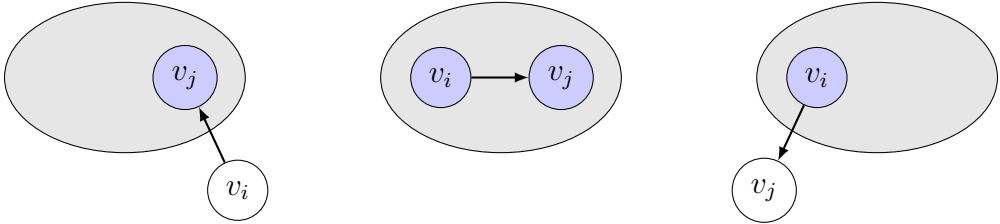

	The PA network model is an evolutionary model assuming that 
	nodes 
	with large degrees are more likely to be connected by new 
	nodes~\citep{Barabasi1999emergence}. We incorporate edge weights 
	into a directed PA network model with five parameters
	$(\alpha, \beta, \gamma, \delta_i,
	\delta_2)$~\citep{Wan2017fitting, yuan2023generating}. 
	Specifically,
	the growth scheme of the extended PA network model 
	is: (1) with probability $0 \le \alpha \le 1$, 
	$(v_i, v_j, w_{ij})$ is added from a new node $v_i$ to an 
	existing 
	node $v_j$; (2) with probability $0 \le \beta \le 1$, 
	$(v_i, v_j, w_{ij})$ is added between two existing nodes~$v_i$ 
	and~$v_j$; (3) 
	with probability $0 \le \gamma \le 1$,  $(v_i, v_j, w_{ij})$ is 
	added from an existing node~$v_i$ to a new node~$v_j$.
	The weight of each edge is independently drawn from a 
	probability distribution~$h$ with support on~$\mathbb{R}^{+}$ or 
	its 
	nonempty subset. Regardless of 
	edge-creation scenario, the probability of sampling an existing 
	node, $v_i$ for instance, as a source (or target) node is 
	proportional to $\outs_i + \delta_1$ (or $\ins_i + \delta_2$). 
	See 
	Figure~\ref{fig:PAscenario} for a graphical illustration.

	\begin{figure}[tbp]
		\centering
		\begin{subfigure}[t]{0.63\textwidth}
			\vskip 0pt
			\includegraphics[width = \textwidth]
			{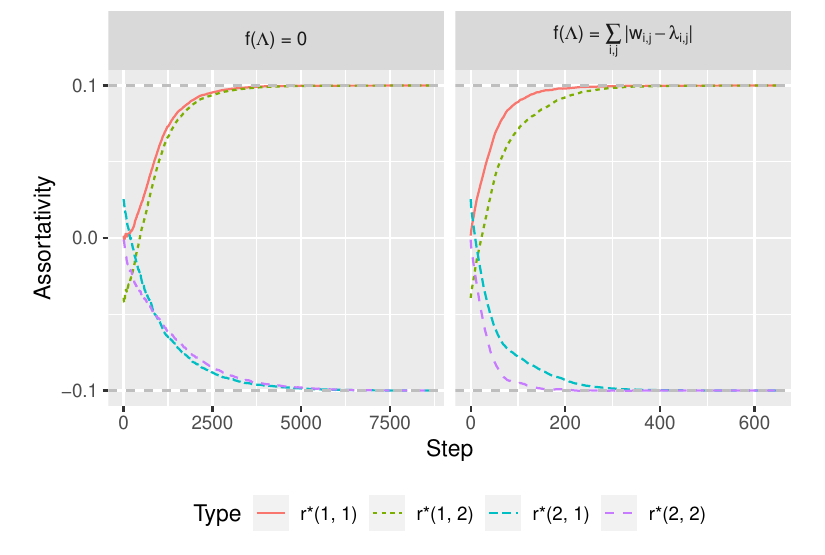}
		\end{subfigure}
		\begin{subfigure}[t]{0.36\textwidth}
			\vskip 0pt
			\includegraphics[width = \textwidth]
			{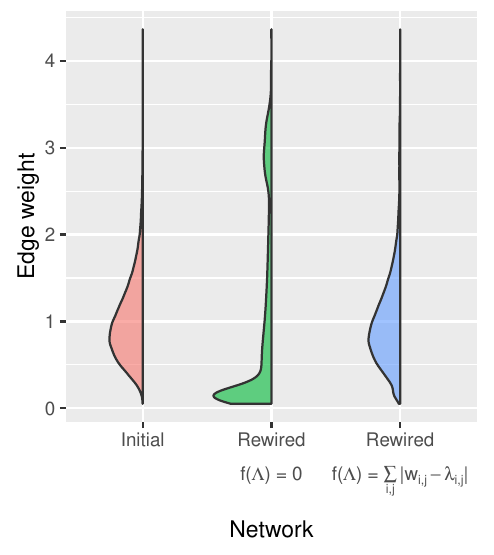}
		\end{subfigure}
		\caption{Average trace and edge weight density plots for PA 
			networks with $m = 1000$, $\alpha = \gamma = 0.15$, 
			$\beta = 0.7$, and target assortativity coefficients 
			$r^*(1, 1) = r^*(1, 2) = 0.1$, $r^*(2, 1) = r^*(2, 2) = 
			-0.1$.
			The left two panels are average trace plots based on 
			objective functions $f(\targetW) = 0$ and 
			$f(\targetW) = \sum_{i, j} |w_{ij} - \lambda_{ij}|$.
			The right panel compares the density of edge weights of 
			the 
			initial networks and the constructed target networks 
			under the same two objective functions.} 
		\label{fig:PA}
	\end{figure}

	The seed network for all of the extended PA networks in our 
	simulation study contained one weighted edge 
	$(1, 2, 1.0)$. 
	The parameters were set to $\beta \in \{0.6, 0.7, 0.8\}$, 
	$\alpha = \gamma = (1 - \beta) / 2$ and $\delta_1 = \delta_2 = 
	1$. 
	Again, $h$ was set to be a gamma distribution with 
	shape~$5$ and scale~$0.2$. We considered PA networks of 
	different 
	sizes determined by number of evolutionary steps 
	$m \in \{200, 400, 600, 800, 1000\}$. 
	The number of replicates for each 
	combination of $m$ and $\beta$ was $100$. The target 
	assortativity 
	coefficients for this series of simulation studies were also
	$r^*(1, 1) = r^*(1, 2) = 0.1$ and $r^*(2, 1) = r^*(2, 2) = -0.1$.

	The PA network simulation study yielded conclusions similar to 
	those
	from the ER network simulation study, despite that the 
	evolutionary
	processes of the two models are tremendously different. 
	Since the results across different $m$ and $\beta$ 
	combinations for PA networks are similar, we only report those 
	for
	$m = 1000$ and $\beta = 0.7$ in Figure~\ref{fig:PA}.
	The left two panels present the 
	average trace plots obtained under different objective functions
	$f(\targetW) = 0$ and 
	$f(\targetW) = \sum_{i, j} |w_{ij} - \lambda_{ij}|$, where the 
	reordering procedure was implemented to both. 
	All the assortativity coefficients reach target values.
	More rewiring steps were needed for objective 
	function $f(\targetW) = 0$, but less computing time was needed. 
	Precisely, the median runtime was $43$ seconds with $f(\targetW) 
	= 0$
	and $142$ seconds for
	$f(\targetW) = \sum_{i, j} |w_{ij} - \lambda_{ij}|$. The
	middle panel shows that more rewiring steps were required for 
	$r^*(1, 2)$ and $r^*(2, 1)$. This was intuitively expected as 
	their 
	initial values were further away from the targets. The right 
	panel
	shows again that the post-rewiring edge weight distribution using
	$f(\targetW) = \sum_{i, j} |w_{ij} - \lambda_{ij}|$ 
	is much closer than that using $f(\targetW) = 0$ to the edge 
	weight
	distribution of the initial networks.

	\begin{figure}[tbp]
		\centering
		\includegraphics[width=0.75\textwidth]{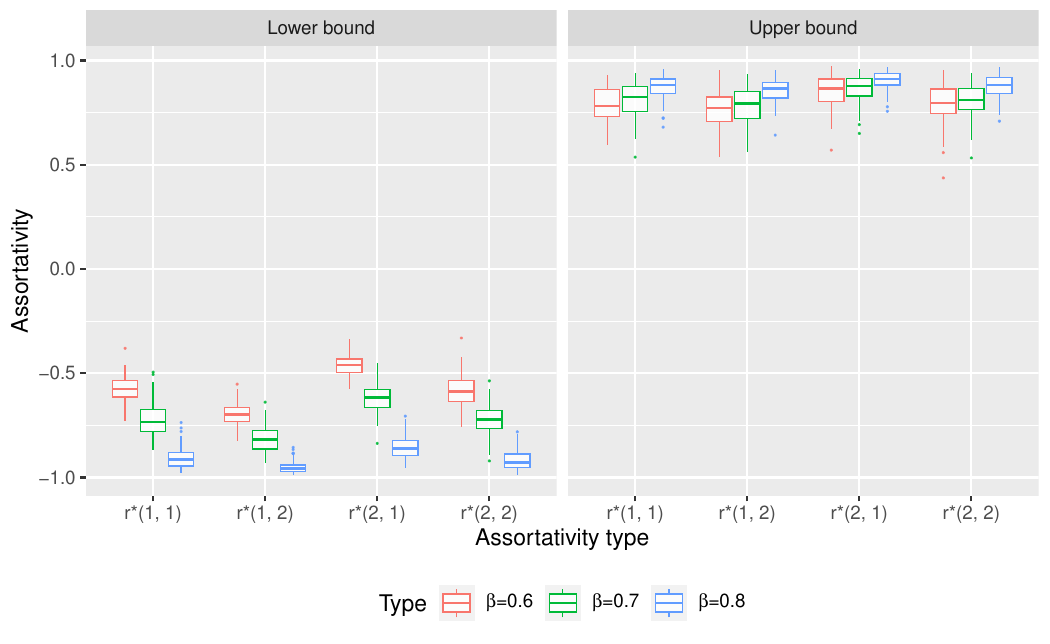}
		\caption{Side-by-side box plots of the upper and lower 
		bounds of 
			the assortativity coefficients of PA networks with 
			$m = 1000$, $\beta \in \{0.6, 0.7, 0.8\}$ and 
			$\alpha = \gamma = (1 - \beta) / 2$.}
		\label{fig:PA_range}
	\end{figure}

	Finally, the box plots of the lower and upper bounds of the 
	attainable
	assortativity coefficients based on 100 replicates for the 
	simulated
	PA networks with the number of evolutionary steps $m = 1000$ are 
	shown in Figure~\ref{fig:PA_range}. With the same value of $m$, 
	the 
	number of edges was fixed, and a larger $\beta$ resulted in a 
	denser 
	network. The range of assortativity coefficients for large 
	$\beta$ 
	was found wider than that for small $\beta$, as larger $\beta$ 
	caused greater variances for node in- and out-strengths.

	\section{Discussions}
	\label{sec:discussion}
	
	The SSRP algorithm tackles the rewiring problem of
	\citet{Newman2003mixing} towards predetermined assortativity 
	levels in
	the context of weighted, directed networks. The rewiring retains 
	the
	certain critical network properties such as the marginal 
	node out- and in-strength distributions and the sparsity.
	The essential idea of the proposed approach is 
	determining the adjacency matrix of a target network by solving a
	mixed integer programming problem, followed by a sweeping
	procedure to transform the initial network to the target by 
	using 
	rewiring. More complex objective functions could be used in 
	setting up
	the mixed integer programming problem at extra computational
	costs to minimize the change in the edge weight distribution.
	The proposed algorithm is also applicable to unweighted or 
	undirected networks with minor modifications.

	There is a major difference between the SSRP algorithm and other
	rewiring methods like Newman's 
	algorithm~\citep{Newman2003mixing} 
	and the DiDPR algorithm~\citep{Wang2022generating}. The SSPR 
	algorithm
	derives an exact solution of target network with predetermined 
	assortativity measures, but the others aim to find an 
	approximated solution, that is, they search a target 
	network with assortativity measures whose expectations equal to 
	the
	given values. Accordingly, the determination of target networks 
	differs between SSPR and other methods, too. SSRP directly works 
	on 
	adjacency matrix calculation, whereas the other methods 
	determine 
	target adjacency matrix through joint node-degree distributions 
	governed by given assortativity measures. This difference means 
	an
	advantage for the SSRP algorithm in some applications and a 
	limitation
	in other applications. Extending the DiDPR algorithm in Newman's 
	sense
	to generating weighted, directed networks remains an open 
	question, yet challenging for preserving network sparsity or 
	other 
	critical network properties in addition to marginal node degree 
	or 
	strength distributions.

	\appendix
	
	\section{Proof of Proposition~\ref{prop:exist}}
	\label{app:proof_exist}
	
	\begin{proof}
		Without loss of generality, assume $\psi_{ij} > 0$. Since we 
		have $\sum_{j = 1}^{n} \psi_{ij} = 0$ according to the 
		rewiring 
		setup, there exists a nonempty set
		$S_j \subseteq \{j + 1, j + 2, \ldots, n\}$
		such that $\psi_{ij^\ast} < 0$ for all
		$j^\ast \in S_j$ and
		$\psi_{ij} + \sum_{j^\ast \in S_j} \psi_{ij^\ast} \leq 0$,
		where the equality holds if $\psi_{ij}$ is the only 
		positive element in the $i$-th row.
		Similarly, due to
		$\sum_{i = 1}^{n} \psi_{ij} = 0$, for each $\psi_{ij^\ast}$ 
		with
		$j^\ast \in S_j$, there exists
		$T_i \subseteq \{i + 1, i + 2, \ldots, n\}$ such that
		$\psi_{i^\ast j^\ast} > 0$ for all
		$i^\ast \in T_i$ and
		$\psi_{ij^\ast} + \sum_{i^\ast \in T_i} \psi_{i^\ast j^\ast} 
		\ge 0$.

		It follows that
		\[
		\sum_{j^\ast \in S_j}\sum_{i^\ast \in T_i}\psi_{i^\ast 
		j^\ast} 
		\ge \psi_{ij},
		\]
		which suggests that, for each pair of $(i^\ast, j^\ast)$, 
		there 
		exists $0 \le u_{i^\ast j^\ast} \le \psi_{i^\ast j^\ast}$ 
		giving rise to
		\[
		\sum_{j^\ast \in S_j}\sum_{i^\ast \in T_i}u_{i^\ast j^\ast}
		= \psi_{ij}.
		\] 
		Therefore, there exists a path continuously rewiring
		$(v_i, v_j, w_{ij})$ and $(v_{i^\ast}, v_{j^\ast}, w_{i^\ast 
		j^\ast})$ with 
		$\Delta w = u_{i^\ast j^\ast}$ for all $i^\ast$ and $j^\ast$ 
		leading to $\psi_{ij} = 0$.

		The proof for $\psi_{ij} < 0$ can be 
		done mutatis mutandis.
	\end{proof}

\end{document}